\documentclass[leqno]{amsart}
\usepackage{a4wide}
\usepackage[usenames]{color}
\usepackage{figsize,subfigure,ifthen,calc,psfrag}
\usepackage{enumerate,comment}
\usepackage{xspace}

\usepackage{latexsym}
\usepackage{amsfonts}
\usepackage{amsmath}
\usepackage{amssymb}
\usepackage{amsxtra}
\usepackage{hyperref}
\usepackage{amsthm}
\usepackage{graphicx}
\usepackage{multicol}
\usepackage{url}
\usepackage{rotating}
\usepackage[bottom]{footmisc}
\usepackage{harvard}
\usepackage{dsfont}

\usepackage{enumitem}
\usepackage{hyperref}
\usepackage{nameref}
\usepackage{hhline}
\usepackage{verbatim}
\numberwithin{equation}{section}

\newcommand{\Cdot}{\raisebox{-0.25ex}{\scalebox{1.3}{$\cdot$}}}

\DeclareMathOperator{\Var}{Var}
\DeclareMathOperator{\Cov}{Cov}
\DeclareMathOperator{\diag}{diag}
\DeclareMathOperator{\vect}{vec}

\begin{document}


\font\srm=cmr8 at 8. truept
\def\bp{{\bar \partial}}
\def\pd{{\partial_\tau}}
\def\bU{{\bar U}}
\def\bW{{\bar W}}
\def\bdel{{\bar{\delta}}}
\def\ubdel{{\underline{\delta}}}
\def\ubk{{\underline k}}
\def\Un2{{U^{n+\frac 12}}}
\def\n2{{n+\frac 12}}
\def\IL{{{I_{\hbox{\srm L}}}}}
\def\Th{{\Theta}}
\def\T_h{{{\mathcal T}_h}}
\def\La{{\Lambda}}
\def\F{{\mathcal{F}}}
\def\G{{\mathcal{G}}}
\def\E{{\mathds{E}}}
\def\H{{\mathcal{H}}}
\def\<{{\langle }}
\def\>{{\rangle }}
\def\V{{\mathcal{V}}}
\def\U{{\mathcal{U}}}
\def\sg{{\sigma_\gamma}}
\def\P{{\mathbf{\mathds P}}}
\def\tends{\rightarrow}

\def\al{\alpha}
\def\f{\varphi}
\def\wU{\widehat U}
\def\wA{\widehat A}
\def\wa{\widehat a}
\def\he{\hat e}
\def\hU{\hat U}
\def\hR{\hat R}
\def\gam{\gamma}
\def\be{\beta}
\def\la{\lambda}
\def\th{\vartheta}
\def\t{\tau}
\def\eps{\varepsilon}
\def\ds{\displaystyle}
\def\C{{\mathbb C}}
\def\Re {{\mathbb R}}
\def\P{{\mathbb P}}
\def\N{{\mathbb N}}
\def\R+{{\Bbb R}_*^+}
\def\Z{{\mathbb Z}}
\def\e{{\rm e}}
\def\i{{\rm i}}
\def\tr|{|\!|\!|}

\def\tm{\tilde{m}}
\def\tp{\tilde{p}}

\def\tf{f}
\newcommand{\ex}{\mbox{E}}
\newcommand{\var}{\mbox{Var}}
\newcommand{\cov}{\mbox{Cov}}
\newcommand{\cum}{\mbox{cum}}
\newcommand{\beps}{\mbox{\boldmath  $\varepsilon $}}
\newcommand{\imag }{\mathrm i}

\newtheorem{theorem}{Theorem}[section]
\newtheorem{lemma}[theorem]{Lemma}
\newtheorem{corollary}[theorem]{Corollary}
\newtheorem{proposition}[theorem]{Proposition}

\newtheorem{remark}[theorem]{Remark}
\newtheorem{example}[theorem]{Example}
\newtheorem{definition}[theorem]{Definition}


\newcommand{\Proof}{{\sc Proof}\, }

\def\Box{\vbox{\offinterlineskip\hrule

        \hbox{\vrule\phantom{o}\vrule}\hrule}}

\newcommand{\bull}{\rule{2mm}{2mm}}

\title[\sf Profiting from correlations]
{\sf Profiting from correlations: Adjusted estimators for categorical data}
\author[\sf T. Niebuhr]{Tobias Niebuhr}
\author[\sf M. Trabs]{Mathias Trabs}
\address{Universit\"at Hamburg, Fachbereich Mathematik, Bundesstrasse 55, D--20146 Hamburg, Germany. email: tobias.niebuhr@uni-hamburg.de; mathias.trabs@uni-hamburg.de}
\date{\today}

\keywords{contingency table; categorical data; empirical frequencies; traffic accident data.} 

\begin{abstract}To take sample biases and skewness in the observations into account, practitioners frequently weight their observations according to some marginal distribution. The present paper demonstrates that such weighting can indeed improve the estimation. Studying contingency tables, estimators for marginal distributions are proposed under the assumption that another marginal is known. It is shown that the weighted estimators have a strictly smaller asymptotic variance whenever the two marginals are correlated. The finite sample performance is illustrated in a simulation study. As an application to traffic accident data the method allows for correcting a well-known bias in the observed injury severity distribution.
\end{abstract}
\maketitle
\allowdisplaybreaks

\section{Introduction}

Categorical data analysis is fundamental for many research fields and applications.
When applying strategies to real-world data, it seems to be the rule rather than the exception that datasets are lacking to some degree. This is usually due to under-reportings of subgroups of the population, to non-representative study participants, or miscoded observations (see e.g. \citeasnoun{Lu2000}, \citeasnoun{Hazell2006}, \citeasnoun{Yamayoto2008} or \citeasnoun{Wang2015}, and the applications therein). All these scenarios have in common that they result in a somehow skewed observation distribution. A major task in applied studies is to address the data's skewness at first before proceeding with any other investigation.

The most common way to correct for such a skewness is the application of weighting factors. However, a weighted or cloned sample does not give any additional information on the underlying random mechanisms, a fact that is often ignored by practitioners.
Statistical methods have to incorporate the weighting step instead of treating the data clones as new and independent observations.
This cloning of dependencies makes the use of weighted samples rather challenging.

In the univariate case, it can be easily seen that weighting or cloning of data can only increase the (asymptotic) variance of distribution estimators compared to an unweighted approach. In higher dimensions the situation turns out to be different.
We will consider a two-dimensional discrete distribution from which one marginal is assumed to be known or can be estimated with higher accuracy. Such additional information in one category is reasonable in many applications, for instance, in our leading example from traffic accident research, where some properties of the underlying population are detailed reported in large databases and statistical surveys, while data on other categories are collected only in specific (and smaller) studies.
The intriguing question is whether the standard unweighted empirical probability estimators can be improved using an additional amount of information.

We will settle this question to the positive.
The constructed estimators are asymptotically unbiased and adapt to the known marginal distribution as desired in practice.
They also satisfy a central limit theorem with a substantially smaller asymptotic covariance matrix (in the sense of positive semi-definite matrices) than the classical estimators. The information theoretic gain is linked to the dependence between the two marginals. 

Our weighted estimation approach is related to the classical problem of ranked contingency tables where a two-way table should be adjusted to the two fixed marginals. Following \citeasnoun{DemingStephan1940}, there is a series of papers on this topic, for instance, \citeasnoun{IrelandKullback1968} and \citeasnoun{LittleWu1991}. However, we have a different target since we aim for the estimation of an unknown marginal instead of the estimation of the contingency table itself.


In addition to our asymptotic analysis, a simulation study reveals a good finite sample performance of the proposed method. Already for small sample sizes considerable improvements for dependent marginals are reported. As a little price to pay, we observe a small finite sample bias. 

Our motivating application comes from traffic accident research. The official national statistics, as, e.g., the National Highway Traffic Safety Administration for the United States or the Statistisches Bundesamt for Germany, provide highly accurate information about the accident situation. However, the administrations usually provide only very low dimensional data. To evaluate new driver assistance systems, the automobile producers require further information on more than one accident characteristic.
Therefore, they retrieve accident studies of more depth, e.g. the National Automotive Sampling System or the German In-Depth Accident Study (GIDAS).
Based on GIDAS data, we estimate the (discretized) distribution of the speed reduction due to collision where the estimators are adjusted for the injury severity distribution from the national statistic. As a result our method allows for correcting a well-known bias in the observed injury severity distribution.

The paper is organized as follows: In Section~\ref{1dim} we briefly discuss the one-dimensional case while the two-dimensional analysis is presented in Section~\ref{2dim}. Section~\ref{numericals} contains the simulation study. The real data example is investigated in Section~\ref{applications}. Section \ref{conclusions} concludes the paper. All technical proofs are deferred to Section~\ref{proofs}.

\section{One-dimensional distributions}\label{1dim}

Before we come to our main results in the next section, we recall some basic facts for the simple univariate case. We observe an independent and identically distributed (i.i.d.)~sample $X_1,\ldots,X_n\in\{1,\dots,I\}$ stemming from a discrete probability distribution $\mathds P =(p_1,\ldots,p_I)$ with $ \sum_{i=1}^I p_i=1 $ and
$$P(X_t=i)=p_i\quad\text{ for all } \, t=1,\ldots,n\,\text{ and } \, i=1,\ldots,I.$$
Here and throughout the $I\in\N$ possible values, or categories, are labeled by $1,\dots,I$ without loss of generality.
The aim is to estimate the probability distribution $\mathds P$.
The usual estimates are computed by the observed relative frequencies
$$\hat p_{i}:=\frac{1}{n}\sum_{t=1}^n \mathds 1_{\{X_t=i\}}\quad\text{ for } i=1,\ldots,I.$$
These estimators satisfy, as $n\rightarrow\infty$, the well-known central limit theorem
\begin{align}\label{CLT}
\sqrt n\big((\hat p_{1},\ldots,\hat p_{I})^\top-(p_{1},\ldots,p_{I})^\top\big)
\stackrel{\mathcal D}{\rightarrow}\mathcal N(0,\Sigma),
\end{align}
where $\stackrel{\mathcal D}{\rightarrow}$ denotes weak convergence and where the asymptotic covariance matrix $\Sigma=(\Sigma_{rs})_{r,s=1,\ldots,I}$ is given by
\begin{align}\label{sigma1dim}
\Sigma_{rs}=\begin{cases}-p_{r}p_{s}&,r\neq s\\ p_{r}(1-p_{r})&,r=s\end{cases}.
\end{align}

The estimators $\hat p_{i}$, $i=1,\ldots,I$, can be interpreted as uniformly weighted (or unweighted) means of the observations.
Changing the weights from $1/n$ to some general weights $w_t\ge0$, $t=1,\ldots,n$, with $\sum_{t=1}^nw_t=1$, yields the more general estimators
\begin{align}\label{tildepi}
\tilde p_{i}:=\sum_{t=1}^n w_t\mathds 1_{\{X_t=i\}},\quad i=1,\ldots,I,
\end{align}
for which we observe a slightly different limiting result whose proof is straightforward:
\begin{lemma}\label{lemmaweighted}
Let $X_1,\ldots,X_n\in\{1,\dots,I\}$ be an i.i.d. sample stemming from a discrete probability distribution $\mathds P =(p_1,\ldots,p_I)$, $ I \in \N $. Then the estimators $(\tilde p_{1},\ldots,\tilde p_{I})^\top$ from \eqref{tildepi} satisfy, as $n\rightarrow\infty$,
\begin{align}\label{weightedIID}
\frac{1}{\sqrt{\sum_{t=1}^n w_t^2}}\big((\tilde p_{1},\ldots,\tilde p_{I})^\top-(p_{1},\ldots,p_{I})^\top\big)
\stackrel{\mathcal D}{\rightarrow}\mathcal N(0,\Sigma),
\end{align}
where $\Sigma$ is given by \eqref{sigma1dim}.
\end{lemma}


By the Cauchy-Schwartz inequality we have $1=\sum_{i=1}^n w_i \leq\sqrt{\sum_{i=1}^n w_i^2}\sqrt{n}$ and thus $\frac{1}{\sqrt{\sum_{i=1}^n w_i^2}}\leq \sqrt n$. Equality holds true if and only if the weights are chosen as $w_t=1/n$ for all $t=1,\ldots,n$. Consequently, in a univariate i.i.d. framework non-uniformly weighting of observations always leads to an increased variance and slower convergence rates resulting in wider confidence interval approximations for finite sample sizes.
We should emphasize at this point that this property might change when leaving the i.i.d. framework and dependencies between the observations are allowed. 

\begin{remark}\label{clone}
In some situations appliers prefer to use data clones instead of (relatively) weighted observations. This may be due to easier implementations when working with databases since it is relatively simple to copy cases. Lemma \ref{lemmaweighted} applies immediately to cloned data which can be seen as follows: Let $Z_t\in\mathds N$ for all $t=1,\ldots,n$, be a deterministic sequence with $\sum_{t=1}^nZ_t=N$. Any observation $X_t$ may be cloned $Z_t$ times such that the sample size artificially increases. The new (cloned) sample then reads $Y_1,\ldots,Y_N$, where $Y_{1+\sum_{k=1}^{t-1}Z_k}=\ldots=Y_{\sum_{k=1}^{t}Z_k}=X_t$ for all $t=1,\ldots,n$. The corresponding estimator for $p_i$ is defined by 
$$
\check p_i=
\frac{1}{N}\sum_{r=1}^N\mathds 1_{\{Y_r=i\}}
=\frac{1}{N}\sum_{t=1}^nZ_t\mathds 1_{\{X_t=i\}}.
$$
We recover the estimators from \eqref{tildepi} with weights $w_t:= \frac{Z_t}{N}$. Since the minimum variance is obtained for non-cloned observations, i.e. $Z_t=1$ for all $t=1,\ldots,n$, we conclude that cloning increases the variability in a one-dimensional model.
\end{remark}


\section{Contingency tables}\label{2dim}

The findings of the previous section directly transfer to the multivariate case. However, we will now investigate whether the situation changes if we have some extra information. Considering a two-dimensional contingency table, we study the estimation of one marginal distribution, say the first, under the assumption that the other marginal distribution, the second, is fully known. 

Let $(X_1,Y_1),\ldots,(X_n,Y_n)$ be a two-dimensional sample taking values in $\{1,\dots,I\}\times\{1,\dots,J\}$ for some $I,J\in\mathds N$ and being distributed according to the law $\mathbf{\mathds P}=(p_{ij})_{i=1,\ldots,I;j=1,\ldots,J}$. Having $I,J$ fixed, we consider a multinomial sampling scheme. The first and second marginals, or equivalently row and column marginal, are denoted by $(p_{1{\Cdot}},\dots,p_{I\Cdot})$ and $(p_{\Cdot 1},\dots,p_{\Cdot J})$, respectively. Without loss of generality, we can assume $p_{\Cdot j}>0$ for all $j=1,\ldots,J$. This setting also includes higher dimensional distributions since we can represent all coordinates with known (joint) distribution as some multi-dimensional random variable $(Y_r)_{r=1,\ldots,R}$, $R\in\mathds{N}$, in a finite state space while the remaining coordinates with unknown marginal distribution can be written as some $(X_s)_{s=1,\ldots,S}$, $S\in\mathds{N}$.

The commonly used estimators for the two-dimensional probabilities $p_{ij}$ are given by
\begin{align*}
\hat p _{i j} = \frac{1}{n}\sum_{t=1}^n \mathds 1_{\{X_t=i,Y_t=j\}}\quad \text{ for all } i=1,\ldots,I,j=1,\ldots,J.
\end{align*}
The resulting estimator for the first marginal distribution is then defined via
\begin{align*}
  \hat p _{i \Cdot} &:= \frac{1}{n}\sum_{t=1}^n \mathds 1_{\{X_t=i\}}=\sum_{j=1}^J\hat p_{ij},\quad  i=1,\ldots,I.
\end{align*}
In view of \eqref{CLT}, these estimators satisfy, as $n\rightarrow\infty$,
\begin{align}\label{CLT2dim}
\sqrt n\left(
(\hat p_{1\Cdot},\ldots,\hat p_{I\Cdot})^\top
-
(p_{1\Cdot},\ldots,p_{I\Cdot})^\top
\right)
\stackrel{\mathcal D}{\rightarrow}
\mathcal N(0,\Sigma)
\end{align}
where $\Sigma=(\Sigma_{rs})_{r,s=1,\ldots,I}$ is given by
\begin{equation}\label{sigma}
\Sigma_{rs}=\begin{cases}p_{r\Cdot}(1-p_{r\Cdot})&,r=s\\ -p_{r\Cdot}p_{s\Cdot}&,r\neq s\end{cases}.
\end{equation}

The estimators $(\hat p_{1\Cdot},\ldots,\hat p_{I\Cdot})$ ignore any specific information on the second components of the sample.
In other words, the estimators treat the two-dimensional sample as a one-dimensional sample. In order to incorporate the additional information to our estimates, we weight the estimators $\hat p_{ij}$ such that the column-wise marginal distribution of the estimators coincide with the true and known marginal distribution $(p_{\Cdot 1},\ldots,p_{\Cdot J})$. More precisely, we introduce the weighted estimators as
\begin{equation}\label{pTilde}
\tilde p_{ij}:=\frac{p_{\Cdot j}}{\hat p_{\Cdot j}}\hat p_{ij}\quad \text{ for any } i=1,\ldots,I;j=1,\ldots,J.
\end{equation}
Owing to the assumption $p_{\Cdot j}>0$, $j=1,\ldots,J$, we have $\hat p_{\Cdot j}>0$ with probability one for a sufficiently large sample size. Hence, due to the above definition we indeed obtain $\tilde p_{\Cdot j}:=\sum_{i=1}^I\tilde p_{ij}=p_{\Cdot j}$ for all $j=1,\ldots,J$. 

Following the discussion in Remark~\ref{clone}, the weighted estimators $\tilde p_{ij}$ can be implemented via careful cloning. For instance, if every entry in the database with $Y_t=j$ is copied $np_{\Cdot j}$ times, then the relative frequencies in the modified dataset coincide with $\tilde p_{ij}$.

\begin{remark}
  The assumption that $\P^Y=(p_{\Cdot 1},\dots,p_{\Cdot J})$ is known can be considerably relaxed. The following results can be extended to the case where the second marginal distribution can be estimated with higher accuracy such that the additional estimation error is negligible, for instance, if we have another i.i.d. sample $Y'_1,\dots,Y'_m\sim \P^Y$ at hand where the sample size $m\in\N$ satisfies $\frac nm\to0$.
\end{remark} 

While the adjusted estimators are fairly simple to implement, they are not anymore linear in the data owing to the data dependent weights. The non-linearity causes a small bias which is negligible compared to the parametric rate $n^{-1/2}$. 
\begin{lemma}\label{lem:bias}
  Let $(X_1,Y_1),\ldots,(X_n,Y_n)\in\{1,\dots,I\}\times\{1,\dots,J\}$ be an i.i.d. sample of a two-dimensional distribution $\mathds P=(p_{ij})_{i=1,\ldots,I;j=1,\ldots,J}$, with some $I,J\in\mathds N$.
Then the estimators $\tilde p_{ij}$ from \eqref{pTilde} satisfy $|\E[\tilde p_{ij}]-p_{ij}|\le 2/n$ for all $1\le i\le I,1\le j\le J$.
\end{lemma}

Using the weighted estimators $\tilde p_{ij}$, the modified contingency table reads 
$$
\begin{array}{ccccc|c}
\tilde p_{11}&&\cdots&&\tilde p_{1J}&\tilde p_{1\Cdot}\\
\vdots&&&&\vdots&\vdots\\
\tilde p_{I1}&&\cdots&&\tilde p_{IJ}&\tilde p_{I\Cdot}\\
\hline
p_{\Cdot 1}&&&&p_{\Cdot J}&1
\end{array}
$$
where the estimates of interest are given by
$$\tilde p_{i\Cdot}:=\sum_{j=1}^J\tilde p_{ij}\quad \text{for all}\quad j=1,\dots,J.$$
It is worth to mention that the adjusted estimators $\tilde p_{ij}$ do not change the cross-product ratios
\begin{align*}
\frac{p_{ij}p_{rs}}{p_{rj}p_{is}},\quad i,r=1,\ldots,I; j,s=1,\ldots,J,\quad
i\neq r,\;j\neq s,
\end{align*}
by construction. These ratios describe the degree of association in a contingency table.
If, for instance, all cross-product ratios in a contingency table are equal to 1, the table yields independence of the marginals, while values of zero or infinity describe dependence.
Since the weighted estimators maintain the two-dimensional inner dependence structure for raw data, cross-product ratio-based tests, e.g. for independence (\citeasnoun{Fisher1962}, \citeasnoun{Goodman1964}), can be applied to the modified contingency table without any restrictions and changes in  interpretation.

\begin{remark}
The choice of the weights is in line with a single iteration of the so-called iterative proportional fitting (IPF) procedure by \citeasnoun{DemingStephan1940}.
By taking marginal distributions of lower dimension, the IPF algorithm computes a higher dimensional distribution fulfilling the given marginal restrictions (cf. \citeasnoun{Stephan1942}). 
The algorithm works conditional on a specific initialization table.
This situation can be interpreted as a two-dimensional sample from which both marginals are known. 
The procedure aims at adapting the sample characteristics to the given marginals.
In the present setup, the weighted sample can be regarded as the result of a single IPF iteration such that the following considerations also hold for any iteration step of the IPF algorithm. For further details on the IPF we refer to \citeasnoun{Fienberg1970} and \citeasnoun{Ruschendorf1995}.
\end{remark}


Intuitively, the inclusion of further information should at least not worsen the estimators' features. To rigorously evaluate whether there is an improvement from the original estimators $(\hat p_{i\Cdot})_{i=1,\ldots, I}$ to the modified estimators $(\tilde p_{i\Cdot})_{i=1,\ldots, I}$, we will compare their limiting variances. 

\begin{theorem}\label{ThTilde}
Let $(X_1,Y_1),\ldots,(X_n,Y_n)\in\{1,\dots,I\}\times\{1,\dots,J\}$ be an i.i.d. sample of a two-dimensional distribution $\mathds P=(p_{ij})_{i=1,\ldots,I;j=1,\ldots,J}$, with some $I,J\in\mathds N$.
Then the estimators 
$\tilde p_{i\Cdot}=\sum_{j=1}^J\tilde p_{ij},i=1,\dots,I,$ with $\tilde p_{ij}$ from \eqref{pTilde} satisfy, as $n\rightarrow\infty$, the limiting result
$$
\sqrt n \big((\tilde p_{1\Cdot},\ldots,\tilde p_{I\Cdot})^\top-(p_{1\Cdot},\ldots,p_{I\Cdot})^\top\big)
\stackrel{\mathcal D}{\rightarrow}
\mathcal N(0,\Gamma),
$$
where $\Gamma=(\Gamma_{k,l})_{k,l=1,\ldots,I}$ is given by
\begin{equation}\label{gamma}
\Gamma_{k,l}=p_{k\Cdot}\mathds 1_{k=l}-\sum_{j=1}^J \frac{p_{k j}p_{l j}}{p_{\Cdot j}}.
\end{equation}
\end{theorem}
The quite simple and explicit structure of the asymptotic variance $\Gamma$ is charming. It can be rigorously understood in terms of the conditional (co-)variance. The latter is defined by
\[
\Cov(V,W|Z):=\E\big[(V-\E[V|Z])(W-\E[W|Z])\big|Z\big]
\]
for arbitrary random variables $V,W$ and $Z$ on the same probability space.
In the following, the notation $\Var(V|Z):=\Cov(V,V|Z)$ also is used. 
As verified by the following lemma, the asymptotic covariance matrix $\Gamma$ of the weighted estimators for the first marginal is given by the covariance of the unweighted estimators conditional on the second component. Hence, we recover in the asymptotic scale that the second margin is known.
\begin{lemma}\label{condCov}
  Grant the assumption of Theorem~\ref{ThTilde}. We then have for any $k,l\in\{1,\dots,I\}$ and $\Gamma_{k,l}$ from \eqref{gamma} that 
  \[
    \frac{1}{n}\Gamma_{k,l}=\E\big[\Cov(\hat p_{k\Cdot},\hat p_{l\Cdot}|Y_1,\dots,Y_n)\big].
  \]
\end{lemma}
This lemma implies together with the law of total variance that for any $i=1,\dots,I$
\begin{equation}\label{eq:ltv}
  \frac{\Sigma_{i,i}}{n}=\Var(\hat p_{i\Cdot})=\E\big[\Var(\hat p_{i\Cdot}|Y_1,\dots,Y_n)\big]+\Var\big(\E[\hat p_{i\Cdot}|Y_1,\dots,Y_n]\big)\ge\frac{\Gamma_{i,i}}{n}.
\end{equation}
Therefore, the asymptotic variance of the adjusted estimators $\tilde p_{i\Cdot}$ is indeed less or equal to the asymptotic variance of the classical estimators $\hat p_{i\Cdot}$. The (asymptotic) information theoretic gain is given by $\Var\big(\E[\hat p_{i\Cdot}|Y_1,\dots,Y_n]\big)$. This fact can be generalized to the joint limit law of $(\tilde p_{1\Cdot},\dots,\tilde p_{I\Cdot})$.

\begin{corollary}\label{ThComparison}
Grant the assumptions of Theorem \ref{ThTilde}. Then the asymptotic covariance matrices $\Sigma$ and $\Gamma$ given in \eqref{sigma} and \eqref{gamma}, respectively, satisfy $\Sigma\ge\Gamma$ in the sense of positive semi-definite matrices: For any vector $c\in\mathds R^I$, we have 
\begin{equation*}
  c^\top(\Sigma-\Gamma)c=\Var\Big(\E\Big[\sum_{i=1}^Ic_i\mathds 1_{\{X_1=i\}}\Big|Y_1\Big]\Big)\ge 0. 
\end{equation*}
In particular, $\Sigma=\Gamma$ holds if and only if $(X_1,\dots,X_n)$ and $(Y_1,\dots,Y_n)$ are independent. 
\end{corollary}

Corollary \ref{ThComparison} shows that the modification of the estimators by the additional information on the known marginal asymptotically improves the estimation of the target marginal. 
The degree of improvement depends on the degree of association of the two marginals. To be more precise, the following bound for the overall relative variance reduction can be easily deduced from \eqref{eq:ltv}:
\begin{align*}
  \sum_{i=1}^I\frac{\Sigma_{ii}-\Gamma_{ii}}{\Sigma_{ii}}
  &\ge \sum_{i=1}^I\frac{\Sigma_{ii}-\Gamma_{ii}}{p_{i\Cdot}}
  = \sum_{i=1}^I\sum_{j=1}^J\frac{(p_{ij}-p_{i\Cdot}p_{\Cdot j})^2}{p_{i\Cdot}p_{\Cdot j}}.
\end{align*}
On the right-hand side we recover a natural measure for dependence being the population counterpart of the well known $\chi^2$-test statistic.

\section{Numerical results}\label{numericals}
Theorem \ref{ThTilde} and Corollary \ref{ThComparison} state that using the modified estimators is asymptotically advantageous over using estimators which ignore the additional information. We have seen that the degree of improvement is related to the degree of association. Focusing first on this effect in a finite sample situation, we investigate the gain of the modification by simulations in 2x2 contingency tables with fixed marginals by varying the degree of association in terms of the cross-product ratio $cpr=(p_{11}p_{22})/(p_{12}p_{21})$ and several sample sizes $n$.

We consider three combinations of marginals, namely
\begin{align*}
\text{(I)}&\quad (p_{1\Cdot},p_{2\Cdot})=(0.5,0.5)\quad \text{and}\quad (p_{\Cdot 1},p_{\Cdot 2})=(0.5,0.5),\\
\text{(II)}&\quad (p_{1\Cdot},p_{2\Cdot})=(0.9,0.1)\quad \text{and}\quad (p_{\Cdot 1},p_{\Cdot 2})=(0.7,0.3),\\
\text{(III)}&\quad (p_{1\Cdot},p_{2\Cdot})=(0.2,0.8)\quad \text{and}\quad (p_{\Cdot 1},p_{\Cdot 2})=(0.7,0.3).
\end{align*}

\begin{figure}[tp]
	\centering
		\includegraphics[width=.5\textwidth]{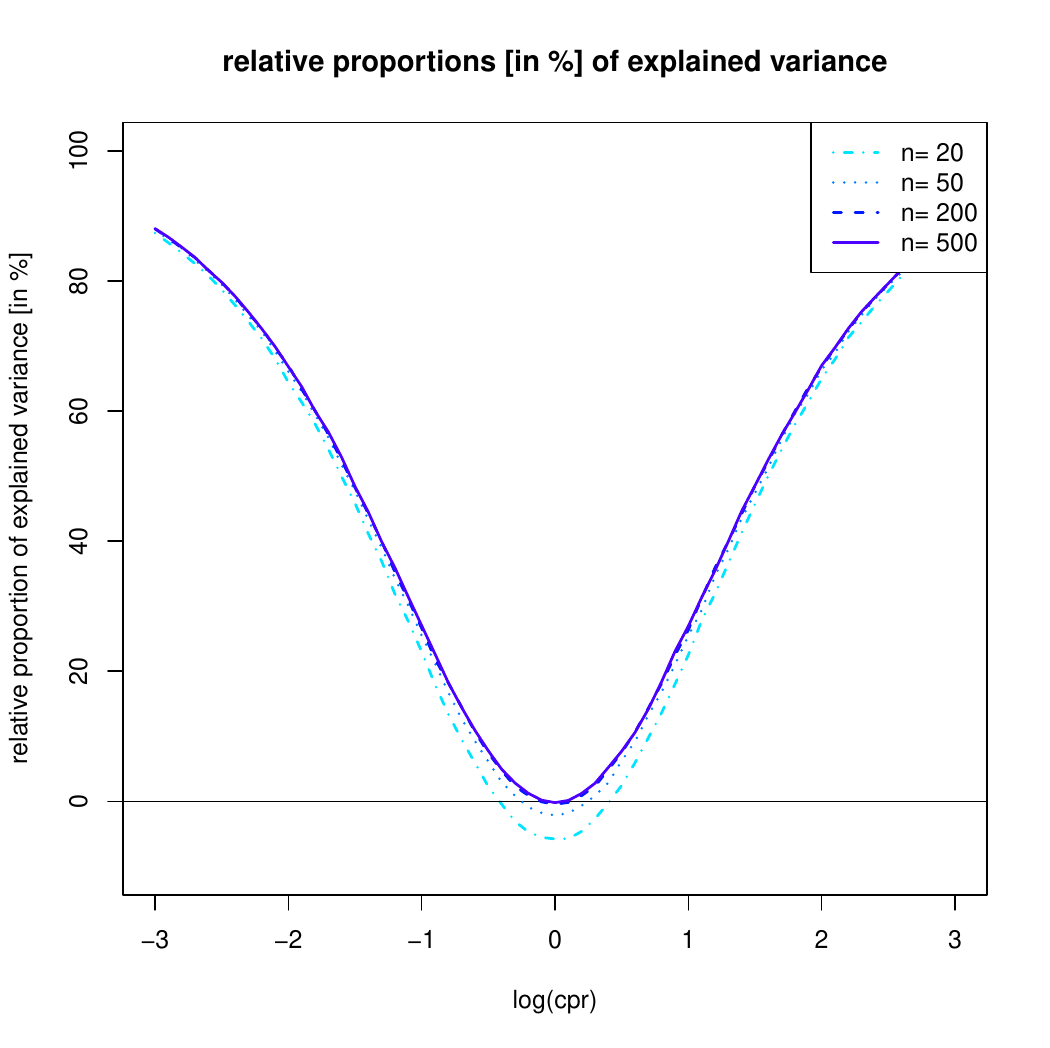}
	\caption{Relative proportions (in \%) of explained variances conditional on the sample size $n$ and the cross-product ratio $cpr$ for the fixed marginals $(p_{1{\displaystyle\cdot}},p_{2{\displaystyle\cdot}})=(0.5,0.5)$ and $(p_{{\displaystyle\cdot} 1},p_{{\displaystyle\cdot} 2})=(0.5,0.5)$.}
	\label{0505}
\end{figure}

\begin{figure}[tp]
	\centering
		\includegraphics[width=.5\textwidth]{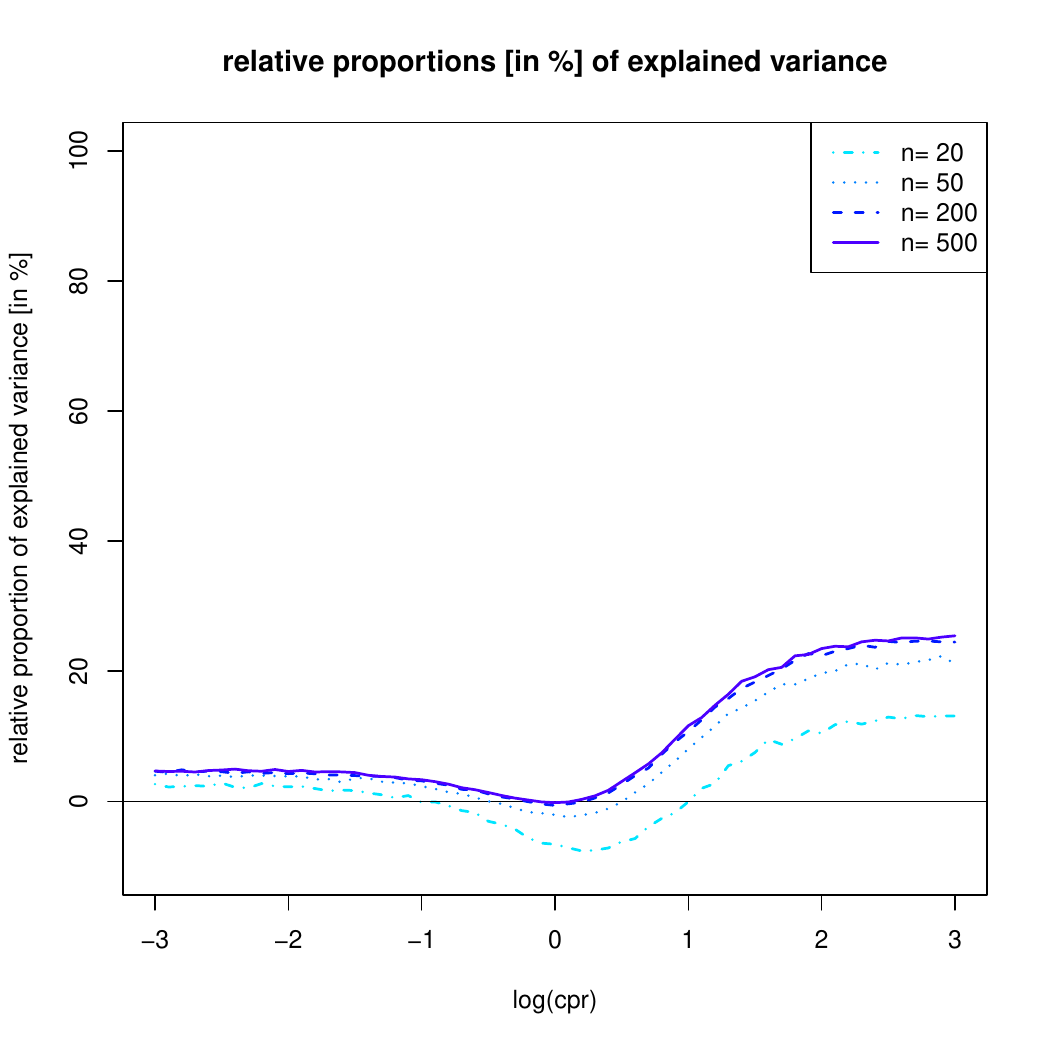}
	\caption{Relative proportions (in \%) of explained variances conditional on the sample size $n$ and the cross-product ratio $cpr$ for the fixed marginals $(p_{1{\displaystyle\cdot}},p_{2{\displaystyle\cdot}})=(0.9,0.1)$ and $(p_{{\displaystyle\cdot} 1},p_{{\displaystyle\cdot} 2})=(0.7,0.3)$.}
	\label{0901}
\end{figure}

\begin{figure}[tp]
	\centering
		\includegraphics[width=.5\textwidth]{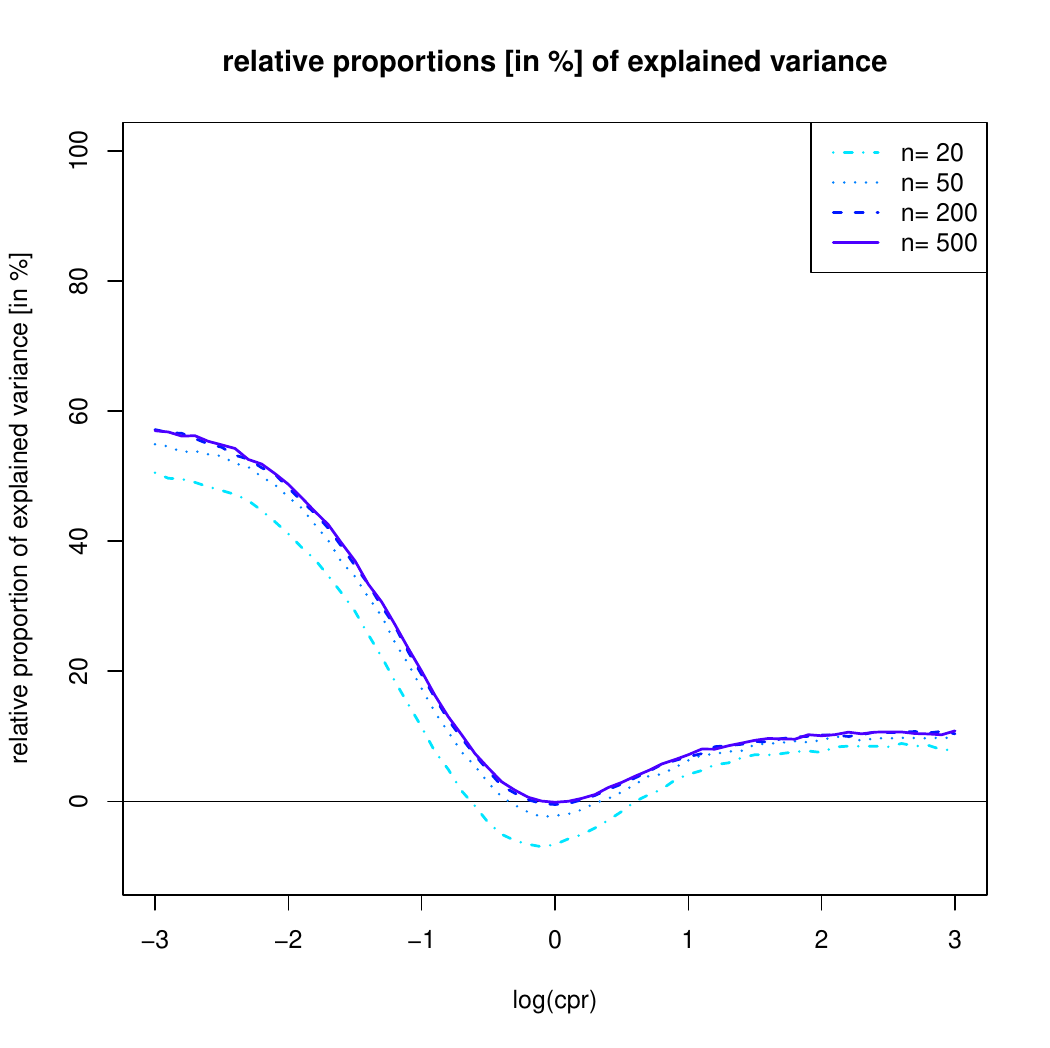}
	\caption{Relative proportions (in \%) of explained variances conditional on the sample size $n$ and the cross-product ratio $cpr$ for the fixed marginals $(p_{1{\displaystyle\cdot}},p_{2{\displaystyle\cdot}})=(0.2,0.8)$ and $(p_{{\displaystyle\cdot} 1},p_{{\displaystyle\cdot} 2})=(0.7,0.3)$.}
	\label{0208}
\end{figure}

Let us focus on the estimation of $p_{1\Cdot}$.
To compare the performance of the estimators $\hat p_{1\Cdot}$ and $\tilde p_{1\Cdot}$, we simulated for any sample size $n\in\{20,50,200,500\}$ and any given cross-product ratio 100,000 samples from which the estimators were computed and noted.
Since the theory has shown that the variance of $\hat p_{1\Cdot}$ should be larger than the variance of $\tilde p_{1\Cdot}$, we denote the average relative proportion of the variance which is removed by the modified estimator $\tilde p_{1\Cdot}$.
The simulation results are given in Figures \ref{0505}-\ref{0208}.
Positive values in the figures indicate whenever the modification of the estimator is -on average- advantageous while negative values indicate that the modification was misleading. Recall that $\log(cpr)=0$ corresponds to independence of the marginals, while the degree of association increases with the distance from the origin in the horizontal axis.


The simulation results show that for already small and moderate sample sizes the modified estimators lead to performance improvements.
It seems that an exception is given in case of independent marginals. In these cases there is a very slight disadvantage for the modified estimator. This effect vanishes when the sample size increases as expected from the theory. Even more important the modified estimators lead to substantial improvements when there is some degree of dependence.

We will now consider a second simulation setup in order to investigate the finite sample performance for large contingency tables. To this end, we generate two independent binomial $Bin(J,p)$-random variables $(Y,Z)$ with success probability $p=0.5$. Setting $X=Y+Z\sim Bin(2J,p)$, we obtain a two dimensional random vector $(X,Y)\in\{0,\dots,2J\}\times \{0,\dots,J\}$ with fixed correlation $\operatorname{Cor}(X,Y)=1/\sqrt 2$ for any dimension $J$. Based on $n$ independent copies of $(X,Y)$, we use the classical as well as the weighted estimators for the marginal distribution $(p_{0,\cdot},\dots,p_{2J,\cdot})$ of $X$. We thus have $I=2J+1$ parameters. Using $100,000$ Monte Carlo iterations, we approximate the $\ell^2$-norm of bias, the overall variance and mean squared error of the estimated vectors $(\hat p_{i,\cdot})_{i=0,\dots, I-1}$ and $(\tilde p_{i,\cdot})_{i=0,\dots, I-1}$. 

The simulation results for $n=1000$ and $J\in\{2,5,10,20,30,50\}$ are reported in Table~\ref{sim}, confirming our theory. We observe a slightly larger bias of the modified estimators for moderate and large dimensions due to the small non-asymptotic bias of $\tilde p_{i\Cdot}$ for any $i=1,\dots,I$. Nevertheless, the squared bias is only of the order of $10^{-4}$ of the variance. The contribution of this bias to the MSE is thus negligible. Consequently, the weighted estimators are in advantage over the classical approach in all considered scenarios. The relative improvement $(\operatorname{MSE}(\hat p)-\operatorname{MSE}(\tilde p))/\operatorname{MSE}(\hat p)$ ranges from 17.2\% for dimension $5$ to 0.7\% for dimension $101$.

\begin{table}
\begin{align*}
\begin{array}{r|ccc|ccc}
I & \operatorname{Bias}(\hat p) & \Var(\hat p) &\operatorname{MSE}(\hat p)&
   \operatorname{Bias}(\tilde p) & \Var(\tilde p) &\operatorname{MSE}(\tilde p)\\
\hline
   5 & 0.731\cdot10^{-4} & 2.459\cdot10^{-4} & 2.459\cdot10^{-4} & 0.691\cdot10^{-4} & 2.035\cdot10^{-4} & 2.035\cdot10^{-4}\\
  11 & 1.008\cdot10^{-4} & 3.188\cdot10^{-4} & 3.188\cdot10^{-4} & 0.984\cdot10^{-4} & 2.895\cdot10^{-4} & 2.895\cdot10^{-4}\\
  21 & 0.879\cdot10^{-4} & 3.651\cdot10^{-4} & 3.652\cdot10^{-4} & 2.257\cdot10^{-4} & 3.449\cdot10^{-4} & 3.449\cdot10^{-4}\\
  41 & 0.815\cdot10^{-4} & 3.991\cdot10^{-4} & 3.991\cdot10^{-4} & 2.934\cdot10^{-4} & 3.869\cdot10^{-4} & 3.869\cdot10^{-4}\\
  61 & 1.205\cdot10^{-4} & 4.184\cdot10^{-4} & 4.118\cdot10^{-4} & 3.460\cdot10^{-4} & 4.104\cdot10^{-4} & 4.105\cdot10^{-4}\\
 101 & 1.258\cdot10^{-4} & 4.354\cdot10^{-4} & 4.354\cdot10^{-4} & 4.348\cdot10^{-4} & 4.322\cdot10^{-4} & 4.323\cdot10^{-4}\\
\end{array}
\end{align*}
  \caption{Monte Carlo approximation of bias $\operatorname{Bias}(\bar p):=\|(\E[\bar p_{i\cdot}]-p_{i\cdot})_i\|_{\ell^2}$, the variance $\Var(\bar p):=\sum_{i=1}^I\Var(\hat p_{i\cdot})$ and the mean squared error $\operatorname{MSE}(\bar p):=\E[\|\bar p_{i\cdot}-p_{i\cdot}\|_{\ell^2}^2]$ of the estimators $(\hat p_{i,\cdot})_{i=0,\dots,I-1}$ and $(\tilde p_{i,\cdot})_{i=0,\dots,I-1}$, respectively.}\label{sim}
\end{table}

\section{A case study from accident research}\label{applications}

The assessment of the effectiveness of advanced driver assistance systems plays a crucial role in traffic accident research.
For reliable analysis detailed information on the pre-crash phase of an accident has to be known to predict possible benefits of future driver assistant systems.
The German In-Depth Accident Study (GIDAS) data contains hundreds of categories which carefully have to be reported for every single accident.
Accidents are reported to GIDAS teams by police, if and only if injured participants are to be expected. This consequently leads to a substantial bias in the injury severity (e.g \citeasnoun{Otte2003}, \citeasnoun{PfeifferSchmidt2006}).
Collision speed or the speed reduction due to the collision, say $\Delta v$, is found to be a major correlate to injury severity in traffic accidents.

In this case study we estimate the (discretized) distribution of $\Delta v$ in passenger car to passenger car collisions with two accident participants.
The corresponding GIDAS subsample contains 8,753 cases.
To address the skewness in the injury severity of the GIDAS sample, we queried the German National Statistic of the year 2014 (\citeasnoun{DESTATIS2016}) from which we found the true injury severity distribution for Germany for these accidents (cf. Table~\ref{global}).

\begin{table}
  \begin{align*}
\begin{array}{l|rrr}
\text{type of injury}&\text{slightly injured}&\text{severely injured}&\text{fatally injured}\\
\hline
\text{number of cases}&	106,181&	11,898&	423\\
&	[89.6\%]&	[10.0\%]&[0.4\%]\\
\end{array}
\end{align*}
\caption{Injury severity distribution in passenger car to passenger car collisions with two accident participants for Germany in 2014 (from Deutsches Bundesamt (2016), p. 100).}\label{global}
\end{table}

Table~\ref{global} can be interpreted as a known marginal distribution for the GIDAS sample.
Hence, having the two-dimensional distribution (about the injury severity and $\Delta v$) from GIDAS at hand, see Table~\ref{gidas}, we can apply the estimation strategy as developed in Section \ref{2dim}.
The estimates for the distribution of $\Delta v$ for the purely GIDAS-based approach, i.e. $\hat p_{i\Cdot}$, are contained in the column \textit{total} of Table~\ref{gidas}.
The National Statistic-aided weighted estimators $\tilde p_{i\Cdot}$ are given in the column \textit{adjusted} besides its relative difference to the purely GIDAS-based estimates.

\begin{table}
\begin{align*}
\begin{array}{lrrr|rr|r|r}
					&\text{slightly}&\text{severely}&\text{fatally}&\text{total}&&\text{adjusted}&\text{relative}\\
					&\text{injured}&\text{injured}&\text{injured}&\text{}&&&\text{difference}\\
\hline
\;\,0\leq \Delta v \leq 10&346&24&2&372		&[11.4\%]&12.0\%&+5.06\%\\
11\leq \Delta v \leq 20&935&118&1&1054	&[32.4\%]&33.6\%&+3.67\%\\
21\leq \Delta v \leq 30&739&192&4&935		&[28.7\%]&28.9\%&+0.43\%\\
31\leq \Delta v \leq 40&335&154&7&496		&[15.2\%]&14.7\%&-3.49\%\\
41\leq \Delta v \leq 50&124&92&7&223		&[6.9\%]&6.3\%&-7.63\%\\
51\leq \Delta v \leq 60&41&50&6&97			&[3.0\%]&2.6\%&-12.36\%\\
61\leq \Delta v \leq 70&16&25&6&47			&[1.4\%]&1.2\%&-15.75\%\\
71\leq \Delta v				 &2&21&7&30				&[0.9\%]&0.7\%&-25.89\%\\
\hline
\text{total}					 &2,538&676&40&3,254&&&\\
											&[78.0\%]&[20.8\%]&[1.2\%]&&[100\%]&&\\
\end{array}
\end{align*}
  \caption{Two-dimensional data pattern (kind of injury and $\Delta v$) from GIDAS in passenger car to passenger car collisions with two accident participants together with marginal estimates for both approaches and their relative difference.}\label{gidas}
\end{table}

From Table~\ref{gidas} we see that the adapted estimates -compared to the ordinary estimates- are increased for lower collision speeds and decreased for higher collision speeds.
On a relative scale, the estimate for the highest $\Delta v$ interval is reduced to around three-quarters of the original estimate.
This is not too surprising.
It was already stated that GIDAS is biased towards the more severely injured traffic participants, cf. \citeasnoun{Otte2003}, or compare Table~\ref{global} to the marginal distribution in Table~\ref{gidas}.
Hence, it is to be expected that accidents with higher collision speeds --which are clearly associated with more severe injuries-- will be down-weighted.
Our proposed method gives precise weights to adjust the more severely injured cases
leading to the final outcome in Table~\ref{gidas}.

\section{Conclusions}\label{conclusions}
The paper investigated how weighting affects the estimation of a discrete probability distribution.
While for one-dimensional data relative weighting will always increase estimation variances, it has been shown that additional information on a marginal distribution in a contingency table allows for estimation improvements of further  marginals if there is some degree of association between the two categories. The gain in terms of the asymptotic (co)variance increases with the degree of dependence. The weighting causes a small bias, which is however negligible compared to the improved variance. For independent marginals the weighted estimators have the same asymptotic behavior as their classical unweighted counterparts. The simulations indicate a clear gain when the marginals are substantially associated. Therefore, from theory and from the simulations perspective we suggest to use the adjusted estimators in applications whenever it cannot be assumed that the marginals are independent. 

\section{Proofs}\label{proofs}

\subsection*{Proof of Lemma~\ref{lem:bias}:}
We decompose
\begin{align*}
  \frac{1}{\hat p_{\Cdot j}}-\frac{1}{p_{\Cdot j}}=\frac{p_{\Cdot j}-\hat p_{\Cdot j}}{p_{\Cdot j}\hat p_{\Cdot j}}
  =\frac{p_{\Cdot j}-\hat p_{\Cdot j}}{p_{\Cdot j}^2}+\frac{(p_{\Cdot j}-\hat p_{\Cdot j})^2}{p_{\Cdot j}^2\hat p_{\Cdot j}}.
\end{align*}
Therefore,
\begin{align*}
 \E\big[\tilde p_{ij}\big]-p_{ij}&=p_{\Cdot j}\E\Big[\Big(\frac{1}{\hat p_{\Cdot j}}-\frac 1{p_{\Cdot j}}\Big)\hat p_{ij}\Big]\\
  &=\frac 1{p_{\cdot j}}\E\big[(p_{\Cdot j}-\hat p_{\Cdot j})\hat p_{ij}\big]+\E\Big[\frac{(p_{\Cdot j}-\hat p_{\Cdot j})^2}{p_{\Cdot j}}\frac{\hat p_{ij}}{\hat p_{\Cdot j}}\Big]
 =:T_1+T_2.
\end{align*}
Since $\hat p_{\Cdot j}$ is unbiased, we obtain for the first term
\begin{align*}
  T_1&=\frac 1{p_{\cdot j}}\E\big[(p_{\Cdot j}-\hat p_{\Cdot j})(\hat p_{ij}-p_{ij})\big]\\
  &\le\frac 1{p_{\cdot j}}\Var(\hat p_{\Cdot j})^{1/2}\Var(\hat p_{i j})^{1/2}\le \frac{1}{n}
\end{align*}
using the Cauchy-Schwarz inequality, $\Var(\hat p_{\Cdot j})=\frac{1}{n}p_{\Cdot j}(1-p_{\Cdot j})$ (and analogously for $\hat p_{ij}$) and $p_{ij}\le p_{\cdot j}$. For the second term the property $\hat p_{ij}\le \hat p_{\cdot j}$ yields
\[
  T_2\le \frac 1{p_{\cdot j}}\Var(\hat p_{\Cdot j})\le\frac1n.
\]
\hfill\fbox\\

\subsection*{Proof of Theorem \ref{ThTilde}:}
We first note that the matrix $(\tilde p_{ij})_{i=1,\ldots,I,j=1,\ldots,J}$ can be understood as a function of all $\hat p_{ij}$. Since a limiting result for the joint distribution of all $\hat p_{ij}$ is given in \eqref{CLT2dim}, the asymptotic features of the joint distribution of the weighted estimators $\tilde p_{ij}$ can be determined using the delta method.

We define the matrices
$$Z=(\tilde p_{ij})_{i=1,\ldots,I;j=1,\ldots,J}\quad\text{and }\quad W=(\hat p_{ij})_{i=1,\ldots,I;j=1,\ldots,J}$$
and calculate
\begin{align*}
Z
&=W\Cdot \diag\left(\frac{p_{\Cdot 1}}{\hat p_{\Cdot 1}},\ldots,\frac{p_{\Cdot J}}{\hat p_{\Cdot J}}\right)\\
&=W\Cdot \diag(\hat p_{\Cdot 1},\ldots,\hat p_{\Cdot J})^{-1} \Cdot \diag(p_{\Cdot 1},\ldots,p_{\Cdot J})\\
&=W\Cdot \diag(\underline 1_I^\top\Cdot W)^{-1} \Cdot \diag(p_{\Cdot 1},\ldots,p_{\Cdot J}),
\end{align*}
where $\underline 1_I\in\mathds R^I$ represents a vector of $I$ repetitions of the value 1.
Now interpret 
\begin{align*}
\left(\begin{array}{c}\tilde p_{1\Cdot}\\\vdots\\\tilde p_{I\Cdot}\end{array}\right)
&=Z\underline 1_J
=W\Cdot \diag(\underline 1_I^\top\Cdot W)^{-1} \Cdot \left(\begin{array}{c}p_{\Cdot 1}\\\vdots\\p_{\Cdot J}\end{array}\right)
\end{align*}
as a function in $W$ or equivalently $\vect(W)=\vect((\hat p_{ij})_{i=1,\ldots,I;j=1,\ldots,J})$.
Hence, $\tilde P:=(\tilde p_{i\Cdot})_{i=1,\ldots,I}$ can be defined by
$\tilde P:\mathds R^{IJ}\rightarrow\mathds R^{I}$, $\vect(W)\mapsto (\tilde p_{i\Cdot})_{i=1,\ldots,I}$.
The entries of the associated Jacobi matrix $J_{\tilde P}\in\mathds R^{I\times (IJ)}$ are computed in the sequel.

To this end, we denote by $\mathds I_{ij}$ the matrix of zeros with a single 1 at the $i$-th row and $j$-th column.
We set $u:=u_{ij}:=i+(j-1)I$ for any $i=1,\ldots,I$ and $j=1,\ldots,J$ and calculate
the $u$-th column of $J_{\tilde P}$ at the point $\vect(W)$ (for brevity we write $\frac{\partial}{\partial W_u}$ instead of $\frac{\partial}{\partial \vect(W)_u}$):
\begin{align*}
&\frac{\partial}{\partial W_{u}}\left(\tilde p_{1\Cdot},\ldots,\tilde p_{I\Cdot}\right)^\top\\
&\quad=\left(\frac{\partial}{\partial W_{u}}W\right)\cdot \diag(\underline 1_I^\top\cdot W)^{-1} \cdot (p_{\Cdot 1},\ldots,p_{\Cdot J})^\top
+ W\cdot \frac{\partial}{\partial W_{u}}\diag(\underline 1_I^\top\cdot W)^{-1} \cdot (p_{\Cdot 1},\ldots,p_{\Cdot J})^\top\\
&\quad= \mathds I_{ij}\cdot \diag(\underline 1_I^\top\cdot W)^{-1} \cdot (p_{\Cdot 1},\ldots,p_{\Cdot J})^\top\\
&\quad\quad+ W\cdot \left(\frac{\partial}{\partial W_{i+(j-1)I}}\diag\left(\frac{1}{(\underline 1_I^\top\cdot W)_1},\ldots,\frac{1}{(\underline 1_I^\top\cdot W)_J}\right)\right)\cdot (p_{\Cdot 1},\ldots,p_{\Cdot J})^\top\\
&\quad= \mathds I_{ij}\cdot \diag(\underline 1_I^\top\cdot W)^{-1} \cdot (p_{\Cdot 1},\ldots,p_{\Cdot J})^\top- W\cdot \diag\left(\left(\frac{\mathds 1_{r=j}}{(\underline 1_I^\top\cdot W)^2_r}\right)_{r=1,\ldots,J}\right)\cdot (p_{\Cdot 1},\ldots,p_{\Cdot J})^\top\\
&\quad= \mathds I_{ij}\cdot \diag(\underline 1_I^\top\cdot W)^{-1} \Cdot (p_{\Cdot 1},\ldots,p_{\Cdot J})^\top
- W\cdot \left(
\frac{p_{\Cdot 1}\mathds 1_{j=1}}{(\underline 1_I^\top\cdot W)_1^2}\,,\dots,\frac{p_{\cdot J}\mathds 1_{j=J}}{(\underline 1_I^\top\cdot W)_J^2}\right)^\top
\\
&\quad= \left(\frac{p_{\Cdot j}\mathds 1_{r=i}}{(\underline 1_I^\top\cdot W)_j}\right)_{r=1,\ldots,I}
- \frac{p_{\Cdot j}}{(\underline 1_I^\top\cdot W)_j^2} (W_{rj})_{r=1,\ldots,I}\\
&\quad= \frac{p_{\Cdot j}}{\hat p_{\Cdot j}}\left(\mathds 1_{i=r}\right)_{r=1,\ldots,I}-\frac{p_{\Cdot j}}{\hat p_{\Cdot j}^2}(W_{rj})_{r=1,\ldots,I}.
\end{align*}
For the delta method to apply, we do not need $J_{\tilde P}$ at $\vect(W)$, but at its non-estimated target value, i.e. we require $J_{\tilde P}$ at $\vect(P):=\vect((p_{ij})_{i=1,\ldots,I;j=1,\ldots,J})$.
By the above computation, the $m$-th row and $(i+(j-1)I)$-th column value is
\begin{align*}
J_{\tilde P,m,i+(j-1)I}(\vect(P))=
-\frac{p_{m,j}}{p_{\Cdot j}}+\mathds 1_{i=m}.
\end{align*}
Hence, 
having all derivatives at hand, the limiting covariance matrix reads
\begin{align*}
\Gamma=J_{\tilde P}\Sigma J_{\tilde P}^\top
\end{align*}
which we can be calculated component-wise to
\begin{align*}
\Gamma_{m,n}
=&\sum_{r=1}^{I}\sum_{s=1}^J\sum_{t=1}^{I}\sum_{v=1}^J J_{\tilde P,m,r+(s-1)I}\Sigma_{r+(s-1)I,t+(v-1)I} J_{\tilde P,n,t+(v-1)I}\\
=&\sum_{r=1}^{I}\sum_{s=1}^J\sum_{t=1}^{I}\sum_{v=1}^J \Big(-\frac{p_{m,s}}{p_{\Cdot s}}+\mathds 1_{r=m}\Big)\left(-p_{r,s}p_{t,v}+p_{r,s}\mathds 1_{r+(s-1)I=t+(v-1)I}\right)
\Big(-\frac{p_{n,v}}{p_{\Cdot v}}+\mathds 1_{t=n}\Big)\\
=&-\sum_{r=1}^{I}\sum_{s=1}^J\sum_{t=1}^{I}\sum_{v=1}^J p_{r,s}p_{t,v}  \Big(-\frac{p_{m,s}}{p_{\Cdot s}}+\mathds 1_{r=m}\Big)\Big(-\frac{p_{n,v}}{p_{\Cdot v}}+\mathds 1_{t=n}\Big)\\
&\hspace{1cm}+ \sum_{r=1}^{I}\sum_{s=1}^J p_{r,s} \Big(-\frac{p_{m,s}}{p_{\Cdot s}}+\mathds 1_{r=m}\Big)\Big(-\frac{p_{n,s}}{p_{\Cdot s}}+\mathds 1_{r=n}\Big)\\
=& -\left(\sum_{r=1}^{I}\sum_{s=1}^J p_{r,s} (-\frac{p_{m,s}}{p_{\Cdot s}}+\mathds 1_{r=m})\right)\left(\sum_{t=1}^{I}\sum_{v=1}^J p_{t,v} (-\frac{p_{n,v}}{p_{\Cdot v}}+\mathds 1_{t=n})\right)\\
&\hspace{1cm}+ \sum_{r=1}^{I}\sum_{s=1}^J \frac{p_{rs}p_{ms}p_{ns}}{p_{\Cdot s}^2}-2\sum_{s=1}^J\frac{p_{ms}p_{ns}}{p_{\Cdot s}}+\sum_{s=1}^Jp_{ns}\mathds 1_{m=n}\\
=& -\left(-\sum_{s=1}^Jp_{ms}+p_{m\Cdot}\right)\left(-\sum_{r=1}^Jp_{nr}+p_{n\Cdot}\right)\\
&\hspace{1cm}+ \sum_{r=1}^{I}\sum_{s=1}^J \frac{p_{rs}p_{ms}p_{ns}}{p_{\Cdot s}^2}-2\sum_{s=1}^J\frac{p_{ms}p_{ns}}{p_{\Cdot s}}+\sum_{s=1}^Jp_{ns}\mathds 1_{m=n}\\
=& \sum_{s=1}^J \frac{p_{ms}p_{ns}}{p_{\Cdot s}}-2\sum_{s=1}^J\frac{p_{ms}p_{ns}}{p_{\Cdot s}}+p_{n\Cdot}\mathds 1_{m=n}\\
=& p_{n\Cdot}\mathds 1_{m=n} - \sum_{s=1}^J \frac{p_{ms}p_{ns}}{p_{\Cdot s}}.
\end{align*}
\hfill\fbox\\

\subsection*{Proof of Lemma~\ref{condCov}}
Since $(X_t,Y_t)_{t=1,\dots,n}$ is an i.i.d. sample, it suffices to consider $n=1$. We have for any $k,l=1,\dots,I$
\begin{align*}
 \Cov(\hat p_{k\Cdot},\hat p_{l\Cdot}|Y_1)
 &=\E\big[\hat p_{k\Cdot}\hat p_{l\Cdot}\big|Y_1\big]-\E\big[\hat p_{k\Cdot}\big|Y_1\big]\E\big[\hat p_{l\Cdot}\big|Y_1\big]\\
 &=\sum_{s=1}^J\sum_{r=1}^J\Big(\E\big[\hat p_{ks}\hat p_{lr}\big|Y_1\big]-\E\big[\hat p_{ks}\big|Y_1\big]\E\big[\hat p_{lr}\big|Y_1\big]\Big).
\end{align*}
Owing to $\hat p_{ks}=\mathds 1_{\{X_1=k\}}\mathds 1_{\{Y_1=s\}}$, we obtain
\begin{align*}
 \Cov(\hat p_{k\Cdot},\hat p_{l\Cdot}|Y_1)
 &=\sum_{s=1}^J\mathds 1_{\{Y_1=s\}}\Big(\E\big[\mathds 1_{\{X_1=k\}}\mathds 1_{\{X_1=l\}}\big|Y_1\big]-\E\big[\mathds 1_{\{X_1=k\}}\big|Y_1\big]\E\big[\mathds 1_{\{X_1=l\}}\big|Y_1\big]\Big)\\
 &=\sum_{s=1}^J\mathds 1_{\{Y_1=s\}}\Big(\frac{p_{k,s}}{p_{\Cdot,s}}\mathds 1_{k=l}-\frac{p_{k,s}p_{l,s}}{p_{\Cdot,s}^2}\Big).
\end{align*}
Therefore,
\begin{align*}
 \E\big[\Cov(\hat p_{k\Cdot},\hat p_{l\Cdot}|Y_1)\big]
 &=p_{k,\Cdot}\mathds 1_{k=l}-\sum_{s=1}^J\frac{p_{k,s}p_{l,s}}{p_{\Cdot,s}}.
\end{align*}
\hfill\fbox\\

\subsection*{Proof of Corollary \ref{ThComparison}:}
In view of \eqref{sigma} and \eqref{gamma} we have
\begin{align*}
\Sigma-\Gamma
=-\left(p_{m\Cdot}p_{n\Cdot}-\sum_{s=1}^J\frac{p_{ms}p_{ns}}{p_{\Cdot s}}\right)_{m,n=1,\ldots,I}.
\end{align*}
We thus obtain for any $c\in\mathds R^I$ that
\begin{align*}
c^\top(\Sigma-\Gamma)c
=&\sum_{i=1}^I\sum_{j=1}^Ic_ic_j
\left(\sum_{s=1}^J\frac{p_{is}p_{js}}{p_{\Cdot s}}-p_{i\Cdot}p_{j\Cdot}\right)\\
=&\sum_{s=1}^J \frac{1}{p_{\Cdot s}}\Big(\sum_{i=1}^I c_ip_{is}\Big)^2-\Big(\sum_{i=1}^I c_ip_{i\Cdot}\Big)^2
=\sum_{s=1}^J p_{\Cdot s}\Big(\sum_{i=1}^I c_i\frac{p_{is}}{p_{\Cdot s}}\Big)^2-\Big(\sum_{i=1}^I c_ip_{i\Cdot}\Big)^2.
\end{align*}
In terms of the function $f_c\colon \{1,\dots,I\}\to \{c_1,\dots,c_I\},i\mapsto c_i$ the last line can be written as
\begin{align}
 c^\top(\Sigma-\Gamma)c&=\E\big[\E[f_c(X_1)|Y_1]^2\big]-\E\big[f_c(X_1)\big]^2\notag\\
 &=\E\big[\E[f_c(X_1)|Y_1]^2\big]-\E\big[\E[f_c(X_1)|Y_1]\big]^2\notag\\
 &=\Var(\E[f_c(X_1)|Y_1])\label{Gam-Sig}\\
 &\ge0,\notag
\end{align}
where equality holds if and only if the conditional expectation $\E[f_c(X_1)|Y_1]$ is constant for any function such function $f_c$. 

It remains to verify the equivalence of $\Sigma=\Gamma$ and the independence of $X_1$ and $Y_1$. Let $X_1$ and $Y_1$ be independent. Then $\E\big[\E[f(X_1)|Y_1]=\E[f(X_1)]$ almost surely for any finite function $f\colon\{1,\dots,I\}\to \mathds R$. Therefore, $c^\top(\Sigma-\Gamma)c=0$ for any $c\in\mathds R^I$, due to \eqref{Gam-Sig}, and thus $\Sigma-\Gamma=0$. For the other direction suppose \eqref{Gam-Sig} is zero for any $c\in\mathds R^I$ and let $i\in\{1,\dots,I\}$ and $j\in\{1,\dots,J\}$ be arbitrary. Choosing $c\in\mathds R^I$ as the vector with 1 in the $i$th row and all other entries zero, we obtain $\E[\mathds 1_{\{X_1=i\}}|Y_1]=P(X_1=i)$ almost surely. We conclude
\[
  P(X_1=i,Y_1=j)=\E\big[\mathds 1_{\{Y_1=j\}}\E[\mathds 1_{\{X_1=i\}}|Y_1]\big]=P(X_1=i)P(Y_1=j)
\]
for all $i=1,\dots,I$ and $j=1,\dots,J$, i.e., $X_1$ and $Y_1$ are independent.
\hfill\fbox\\

\section*{Acknowledgements}
We are indebted to Dr. Mirko Junge and the Volkswagen AG for providing the traffic accident data and for fruitful discussions on the case study. The authors also thank two anonymous referees for helpful comments.



\end{document}